\documentclass[aps,prd,preprint,a4paper,showpacs,showkeys,superscriptaddress]{revtex4}

\usepackage{latexsym,amsmath,amssymb}
\allowdisplaybreaks

\begin{document}


\title{Complementary role of the pressure in the black hole thermodynamics}

\author{Edwin J. Son}
\email[]{eddy@nims.re.kr}
\affiliation{Division of Computational Mathematics,
National Institute for Mathematical Sciences, Daejeon 305-390, Republic of Korea}

\author{Wontae Kim}
\email[]{wtkim@sogang.ac.kr}
\affiliation{Department of Physics, Sogang University, Seoul 121-742, Republic of Korea}
\affiliation{Center for Quantum Spacetime, Sogang University, Seoul 121-742, Republic of Korea}

\date{\today}

\begin{abstract}
In black hole thermodynamics of certain models,
the thermodynamic first law may contain the pressure term. The corresponding entropy follows the 
area law whereas the thermodynamic energy is not the same with the black hole mass.
If the pressure can be decomposed into two parts and recombined with the original thermodynamic quantities,
then the thermodynamic energy becomes the black hole mass and the entropy satisfying the area law 
turns out to be the corrected entropy called the Wald entropy, respectively.
\end{abstract}



\maketitle

\newcommand{\lp}{\ell_P}
\newcommand{\gn}{G_N}

Since the thermodynamic first law for black holes has been extensively used in
black hole thermodynamics and phase transitions~\cite{Bekenstein:1973ur,Bekenstein:1974ax,Hawking:1974sw,Hawking:1982dh}, 
it has been claimed that the first law is essentially related to the Einstein
equation~\cite{Jacobson:1995ab}. Recently, the remarkable similarity between them has been
emphasized by investigating it at the horizon~\cite{Padmanabhan:2012gx}. 
Explicitly, it shows that the Einstein tensor $G_{~r}^r$ for spherically symmetric static
black holes 
can be decomposed into two parts; the first part is of relevance to the
energy, while the second part is associated with the entropy as
long as the Hawking temperature is identified with the surface gravity. 
Additionally, if there is a source terms for matter, 
then the pressure of $P=T_{~r}^r$ from energy-momentum tensors is included in the thermodynamic first law.
After all, it becomes the thermodynamic first law, $dE -T_H dS=-PdV$, where $T_H$ is the Hawking temperature
and $V$ is the volume of the black hole. Subsequently,
we can easily see that what the energy $E$ and the entropy $S$ for the black hole are from 
this relation. 
Furthermore, the relationship between the first law of thermodynamics and gravitational 
field equations in a spherically symmetric space-time has been extensively investigated 
in the modified gravity theories so that the thermodynamic quantities obtained from 
the gravitational field equations are consistent with those from other 
approaches~\cite{Paranjape:2006ca,Akbar:2006mq,Cai:2009ph}.
In particular, the thermodynamic energy of a spherically symmetric black hole in a Lorentz 
noninvariant model has shown to be exactly same with a generalization of the Misner-Sharp 
energy~\cite{Cai:2009ph}.

However, the pressure can be sometimes regarded as a part of the geometric 
part such as the Einstein tensor instead of the source.
It means that the pressure
can be split into two parts and recombined with the original energy and the entropy,
which yield modified ones.
There are at least two kinds of representations depending on whether 
the pressure is kept in the right-hand side of the equation or not.
In this paper, we would like to apply the above two different approaches to 
two special models so that we show that the thermodynamic energy can become the black
hole mass and the entropy can turn out to be the Wald entropy~\cite{Wald:1993nt} if the pressure
is decomposed and recombined with the original energy and the pressure. 

The first illustration is the four-dimensional exact soluble model~\cite{Cai:2009ua}
 in the semiclassical Einstein equation with the conformal anomaly~\cite{Deser:1993yx,Duff:1993wm},
\begin{equation}
\label{eom}
  G_{\mu\nu}= 8\pi \gn \langle T_{\mu\nu} \rangle,
\end{equation}
where $G_{\mu\nu}=R_{\mu\nu} - \frac12 g_{\mu\nu} R $ and 
$\langle T_{\mu\nu} \rangle$ is the effective energy-momentum tensor from
the trace anomaly, 
$g^{\mu\nu} \langle T_{\mu\nu} \rangle = -\alpha (8 \pi \gn)^{-1} E_{(4)} + \tilde{\alpha}
 C_{\alpha\beta\kappa\lambda} C^{\alpha\beta\kappa\lambda},$
where the coefficients $\alpha $ and $\tilde{\alpha}$ depend on matter contents,
$E_{(4)} = R_{\mu\nu\kappa\lambda} R^{\mu\nu\kappa\lambda} - 4 R_{\mu\nu} R^{\mu\nu} + R^2$
is the Gauss-Bonnet term, and $C_{\alpha\beta\kappa\lambda}$ is the Weyl tensor.
By assuming $\tilde{\alpha}=0$ and $\alpha >0$ for the exact solubility with
the relations of $\langle T^t_{~t} \rangle = \langle T^r_{~r} \rangle$ in the whole space-time~\cite{Cai:2009ua},  one can find the asymptotic flat exact solution,
$ds^2 = -f(r) dt^2 + f^{-1}(r) dr^2 + r^2 d\Omega_2^2$,
\begin{equation}
\label{sol:f}
f(r) = 1 - \frac{r^2}{4\alpha} \left[ 1 - \sqrt{1 - \frac{16\alpha \gn M}{r^3} - \frac{4\alpha a^2}{r^4}} \right],
\end{equation}
where $M$ and $a$ are integration constants.
The metric~\eqref{sol:f} has two roots satisfying $f(r_\pm)=0$; however, the 
horizon is obtained as $r_+ = \gn M  +\sqrt{\gn^2 M^2+2\alpha}$ because
$r_-$ becomes negative for $\alpha>0$.
Following the procedure in Ref.~\cite{Padmanabhan:2012gx}, the relevant tensor component for the static spherically symmetric solution~\eqref{sol:f} is given 
by only the radial part,
\begin{equation}
\label{radial equation}
G^r_{~r}=8\pi \gn \langle T^r_{~r} \rangle,
\end{equation}
where the Einstein tensor and the anomaly term can be written as
$G^r_{~r}(r)=-r^{-2}(1-f-rf')$,
$\langle T^r_{~r} \rangle = \alpha (4\pi \gn)^{-1} r^{-4}(1-f)(1-f+2rf')$,
respectively. The prime means the derivative with respect to
the radial coordinate.
Considering  Eq.~\eqref{radial equation} at the event horizon of $r = r_{+}$,
one can obtain the equation of motion,
\begin{equation}
\label{crude first law}
d \left(\frac{r_{+}}{2 \gn}\right)-\frac{f'(r_{+})}{4\pi} d \left(\frac{\pi r_+^2}{4 \gn}\right)=-\langle T_{~r}^r \rangle dV |_{r=r_+},
\end{equation}
where $V$ is the volume of the black hole.
Now, the Hawking temperature is defined by the surface gravity of $T_{\rm H}=f'(r_{+})/4\pi$ and the pressure is given by $P=\langle T_{~r}^r \rangle$.
Requiring the thermodynamic first law,
\begin{equation}
d\tilde{E} - T_H d\tilde{S}=-PdV |_{r=r_+},
\end{equation}
we obtain the thermodynamic energy and the entropy as
\begin{align}
\tilde{E} = \frac{r_+}{2 \gn}, \qquad \tilde{S} = \frac{\pi r_+^2}{\gn} = \frac{A}{4 \gn},
\end{align}
where $A=4\pi r_{+}^2$. Note that the thermodynamic energy is not the same with black hole mass,
while the entropy satisfies the well-known area law in the presence of pressure.
On the other hand, the above quantum-mechanically induced pressure can
be decomposed into two parts and they are absorbed into the energy and the entropy,
so that Eq.~\eqref{crude first law} can be written as
\begin{equation}
\label{new one}
d \left( \frac{r_{+}}{2 \gn}-\frac{\alpha}{\gn r_{+}} \right)-\frac{f'(r_{+})}{4\pi} d \left( \frac{A}{4 \gn}-\frac{4\pi\alpha}{\gn} \ln \frac{A}{A_0} \right)=0.
\end{equation}
Note that there appears the $\alpha$ correction at each term
due to the anomaly.
Comparing Eq.~\eqref{new one} with the thermodynamic first law
of $dE-T_{\rm H }dS=0$,
we can identify the modified thermodynamic energy and the modified entropy as
\begin{equation}
E=M,~~~~~S = \frac{A}{4 \gn} - \frac{4 \pi \alpha}{\gn} \ln \frac{A}{A_0}.
\end{equation}
Note that the energy is nothing but the black hole mass 
which can be compactly written in terms of the inverse relation 
of the event horizon, $M=r_{+}/2-\alpha/r_{+}$, and
the constant $A_0$ reflects the logarithmic ambiguity. 
The entropy receives the logarithmic correction which 
is the same with the Wald entropy in Ref.~\cite{Cai:2009ua}
(another approach can be found in Ref.~\cite{Banerjee:2008fz}). 

Second, let us start with the $d$-dimensional Einstein-Hilbert action $I_\text{EH}$ coupled to a Gauss-Bonnet term $I_1$
with a negative cosmological constant $\Lambda = -(d-1)(d-2)/2\ell^2$,
\begin{equation}
\label{action}
\begin{aligned}
I &= I_\text{EH} + I_{1} \\
 &=\frac{1}{16\pi \gn} \int d^dx \sqrt{-g} R  + \frac{1}{16\pi \gn} \int d^dx \sqrt{-g}\left[ - 2 \Lambda + \frac{\alpha}{(d-3)(d-4)} E_{(4)} \right],
\end{aligned}
\end{equation}
where $d$ is considered greater than four.
Variation of the action~\eqref{action} yields the 
equations of motion,
\begin{equation}
\label{eom:cov}
\begin{aligned}
G_{\mu\nu} &= 8\pi \gn T_{\mu\nu} \\
&\begin{aligned} \equiv -\Lambda g_{\mu\nu} - \frac{\alpha}{(d-3)(d-4)} \bigg[ 2 R R_{\mu\nu} & - 4 R_{\mu\kappa} R^{\kappa}_{\nu} - 4 R^{\kappa\lambda} R_{\mu\kappa\nu\lambda} \\
  & + R_{\mu\kappa\lambda\sigma} R_{\nu}^{~\kappa\lambda\sigma} - \frac12 {g}_{\mu\nu} E_{(4)} \bigg], \end{aligned}
\end{aligned}
\end{equation}
where $ T_{\mu\nu}=-(8\pi \gn)^{-1}\delta I_1 /\delta {g}^{\mu\nu}$.
The metric is assumed to be the static metric of
$ds^2 = -f(r) dt^2 + f^{-1}(r) dr^2 + r^2 d\Omega_{(d-2),k}^2$,
where $d\Omega_{(d-2),k}^2$ is the line element of a $(d-2)$-dimensional totally symmetric space with a constant curvature $(d-2)(d-3)k$.
Then, the explicit form of the solution is given as~\cite{Cai:2001dz},
\begin{equation}
f(r) = k + \frac{r^2}{2\alpha} \left[ 1 \pm \sqrt{1+\frac{64\pi \gn \alpha M}{(d-2)\Sigma_k r^{d-1}}-\frac{4\alpha^2}{\ell^2}} \right],
\end{equation}
where $\Sigma_k$ is the volume of the $(d-2)$-dimensional space.
There are two branches so that we take the ($-$) sign which goes to the Schwarzschild metric asymptotically.
Then the mass can be written in terms of the horizon $r_+$ as 
\begin{equation}
\label{M}
M = \frac{(d-2)\Sigma_k r_+^{d-3}}{16\pi \gn} \left[ k + \frac{r_+^2}{\ell^2} + \frac{\alpha k^2}{r_+^2} \right],
\end{equation}
where the horizon satisfies $f(r_+)=0$.
Using the metric, the relevant Einstein tensor and 
the pressure from the energy-momentum tensors
can be calculated as  
\begin{align}
G_{~r}^r  &= \frac{(d-2)}{2} \left[ \frac{f'(r)}{r} - \frac{(d-3)(k-f(r))}{r^2} \right], \\
{T}_{~r}^r & = -\frac{(d-2)\alpha}{8\pi \gn} \left[ \frac{f'(r)(k-f(r))}{r^3} + \frac{(d-5)}{2r^4} \left( f(r)(2k-f(r)) - k^2 \right) \right] 
 + \frac{(d-1)(d-2)}{16\pi \gn \ell^2}  \label{pressure},
\end{align}
respectively, 
so that the Einstein equation~\eqref{eom:cov} can be written as 
\begin{equation}
\frac{(d-2)(d-3)\Sigma_k k r_+^{d-4}}{16\pi \gn}  dr_+ - \frac{f'(r_+)}{4\pi} \left[ \frac{(d-2)\Sigma_k r_+^{d-3}}{4\gn}\right] dr_+   =-T_{~r}^r  dV |_{r=r_+}.
\end{equation}
Defining the pressure $P=T_{~r}^r $
and the differential volume $dV = \Sigma_k r^{d-2} dr$,  
we can regard the Einstein equation with matter as
 the thermodynamic first law~\cite{Padmanabhan:2012gx},
\begin{equation}
d\tilde{E} - T_H d\tilde{S}=-PdV |_{r=r_+},
\end{equation}
as long as we identify the energy and the entropy with
\begin{equation}
 \tilde{E} = \frac{(d-2) \Sigma_k k r_+^{d-3}}{16\pi \gn}, \qquad
 \tilde{S} = \frac{A}{4\gn},
 \end{equation}
where $A=\Sigma_k r_+^{d-2}$. Note that the thermodynamic energy cannot be written as the black hole mass; 
in particular, the entropy satisfies the area law. 
Similar to the first illustration, the pressure in Eq.~\eqref{pressure}
is transposed to the left-hand side of the Einstein equation and then it 
can be decomposed into two terms as in the following:
\begin{equation}
\label{split}
\begin{aligned}
\frac{(d-2)\Sigma_k}{16\pi \gn} \left[ (d-3) k r_+^{d-4} + \frac{(d-1) r_+^{d-2}}{\ell^2} + (d-5) \alpha k^2 r_+^{d-6} \right] dr_+ & \\
- \frac{f'(r_+)}{4\pi} \left[ \frac{(d-2)\Sigma_k r_+^{d-3}}{4\gn} \left( 1 + \frac{2\alpha k}{r_+^2} \right) \right] dr_+ & = 0 .  
\end{aligned}
\end{equation}
Identifying Eq.~\eqref{split} with the thermodynamic first law of
$ dE = T_H dS$ without the pressure, we can get the modified energy and the 
modified entropy,
\begin{align}
E=M, \qquad
S=\frac{\Sigma_k r_+^{d-2}}{4\gn} \left[ 1 + \frac{2(d-2)\alpha k}{(d-4) r_+^2} \right].
\end{align}
The superficially complicated thermodynamic energy becomes just the black hole mass and the entropy can be written as
the Wald entropy which is compatible with the previous result in Ref.~\cite{Cai:2001dz}.
In particular, as seen from Eq.~\eqref{split}, there is no explicit Gauss-Bonnet correction to the energy in 
five dimensions, whereas the mass formula~\eqref{M} still has an $\alpha$ correction.
However, the $\alpha$ correction term in the mass formula is actually independent of the horizon, so that it 
can be an integration constant.
For a spherically symmetric black hole without the cosmological constant, these relations 
have been obtained in Ref.~\cite{Paranjape:2006ca} and are consistent with our results.

In conclusion,
we have studied two kinds of expressions of the thermodynamic first law 
depending on the existence of the pressure throughout the two black hole models. 
In the presence of the pressure in the thermodynamic law, the entropy of the black hole
can be written as the well-known area law, while the thermodynamic energy cannot be written
as the black hole mass. 
In contrast to this expression,  if the pressure is eliminated, then 
the entropy receives some corrections to the
standard area law, while the thermodynamic energy of the black hole becomes the 
black hole mass. So, the black hole mass and 
the Bekenstein-Hawking entropy seems to be complementary in the presence of the
pressure.

\acknowledgments 
This work was supported by the National Research Foundation of Korea(NRF) grant funded by the Korea government (MEST) (2010-0008359).


\end{document}